# ESO–SKAO Synergies for the Epoch of Reionisation and Cosmic Dawn

Andrei Mesinger[1]
Benedetta Ciardi[2]
James E. Davies[1]
Samuel Gagnon-Hartman[1]
Valentina D'Odorico[3,1]

[1] Scuola Normale Superiore, Pisa, Italy
[2] Max Planck Institute for Astrophysics, Garching, Germany
[3] INAF–Trieste Astronomical Observatory, Italy

**Mapping out the first billion years using the 21-cm line with the Square Kilometre Array (SKA) will revolutionise our understanding of the cosmic dawn, reionisation and the galaxies that drove these milestones. However, synergies with other telescopes in the form of cross-correlations will be fundamental in making and confirming initial, low signal-to-noise claims of a detection. Participants in the 2023 ESO–SKAO workshop discussed such synergies for Epoch of Reionisation (EoR) and Cosmic Dawn (CD) science. Here we highlight some of the most promising candidates for cross-correlating SKA EoR/CD observations with ESO instruments such as the Multi-Object Optical and Near-infrared Spectrograph (MOONS), the MOSAIC multi-object spectrograph, and the ArmazoNes high Dispersion Echelle Spectrograph (ANDES).**

## Introduction

The final phase-change of our Universe, the so-called Epoch of Reionisation (EoR), remains at the forefront of modern cosmology. After decades of studies, we are starting to home-in on the timing of the bulk of reionisation (for example, Qin et al., 2021; Bosman et al., 2022). However, we still do not really know how to connect this final phase-change of our Universe to the populations of stars and black holes that drive it.

The ultimate probe of this fundamental milestone is arguably the 21-cm line of neutral hydrogen, which is sensitive to the ionisation, temperature and density fluctuations of the intergalactic medium (IGM). Current 21-cm interferometers are aiming for a statistical detection of the 21-cm power spectrum (for example, Mertens et al., 2020; Trott et al., 2020; Abdurashidova et al., 2022). However, over the next decade(s), the Square Kilometre Array being built in Australia (SKA-Low) should provide the ultimate dataset: a 3D map of the barely-explored first half of our observable Universe. Such a worthy dataset contains precious insights into the astrophysics of galaxies and the IGM, as well as physical cosmology (see, for example, the review by Mesinger, 2020).

However, it will be a long time until we have a high-signal-to-noise (S/N) map of the EoR and the preceding Cosmic Dawn (CD). Initial claims of a detection will come from a handful of power-spectrum wave modes with low S/N. Given the novel, ground-breaking nature of the observation, it will be challenging to convince ourselves and the broader community that these preliminary 'detections' are genuinely cosmological. The best way of doing this is to cross-correlate the 21-cm signal with another signal of known cosmic origin.

In addition to providing an invaluable sanity check on preliminary claims of a 21-cm detection, cross-correlations can also improve the S/N. The cross power spectrum could provide a cleaner probe of the cosmological signal, since the foregrounds and systematics of different datasets are typically not correlated (for example, Amiri et al., 2024). Eventually, with SKA phase 2 we should be able to correlate images (i.e. including phases) of different datasets. This would allow us to study individual ionised or heated regions, comparing their tomography (obtained with SKA) to the brightest galaxies they contain (obtained with optical/IR telescopes like JWST).

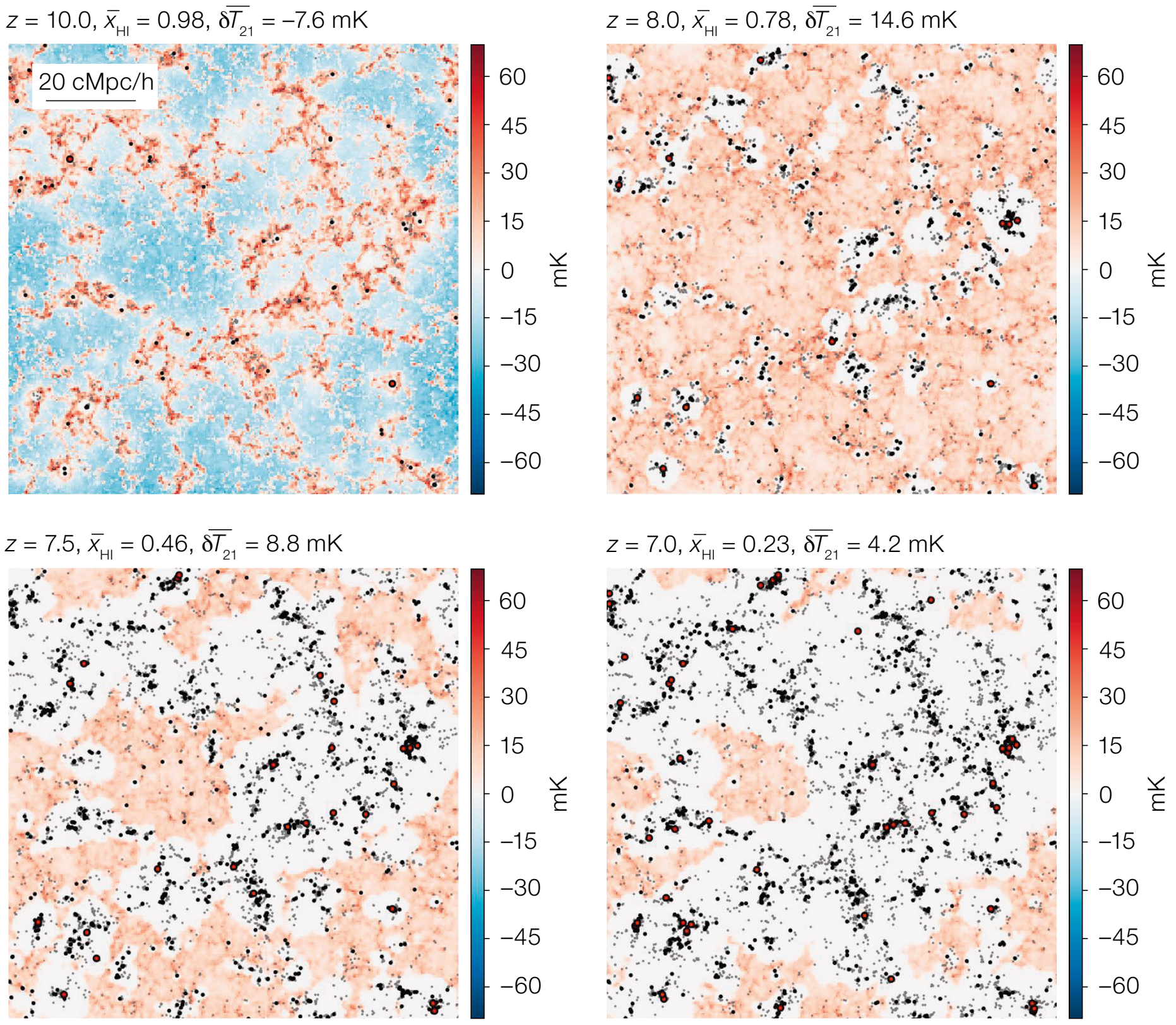

**Figure 1.** An example 21-cm brightness temperature map (see colour bar) and corresponding galaxy map at $z$ = 10, 8, 7.5 and 7, obtained from a 3D multi-frequency radiative transfer code (Eide et al., 2020). The black and red points represent galaxies with [OIII] luminosity $L_{OIII} > 10^{41}$ erg s$^{-1}$ and $L_{OIII} > 10^{42}$ erg s$^{-1}$, respectively, while grey points denote all galaxies within the slice. Galaxy locations are clearly seen to correlate with the 21-cm brightness temperature, since their radiation ionises and heats the surrounding IGM. Figure from Moriwaki et al. (2019).

## EoR signals that can be cross-correlated with the 21-cm line

The 21-cm signal from the EoR/CD is determined by the IGM density, temperature, and ionised fraction. During the dark ages, it is a fairly clean probe of density fluctuations. After the CD, radiation from the first galaxies heats and ionises the IGM, effectively coupling modes over a range of scales. However, galaxies themselves are biased tracers of the matter field. Therefore any tracer of the large-scale matter field should in principle correlate with the 21-cm signal from the EoR/CD. The quantitative details of this correlation encode a wealth of information about the first galaxies (see references below).

In practice, however, it is difficult to cross-correlate signals because we need to match the ‘footprints’ of different probes. Foregrounds for 21-cm interferometry live in a ‘wedge’ region in Fourier space (for example, Morales et al., 2012). As such, large-scale and transverse (on-sky) modes might either be lost during foreground cleaning or be subject to a very large cosmic variance in the cross-correlation (i.e. uncorrelated cross terms only vanish in the limit of infinite samples). Therefore, the ideal candidates for cross-correlation with 21-cm interferometry are large-volume surveys with good redshift localisation.

We list some potential candidates here, briefly mentioning their pros and cons.

1. Cosmic backgrounds: (integral) radiation backgrounds like the cosmic microwave background (CMB; for example, Ma et al., 2018a; La Plante, Sipple & Lidz, 2022), near-infrared background (NIR; for example, Mao, 2014), and the X-ray background (XRB; Ma et al., 2018b), contain a contribution from $z > 5.5$, which should correlate with the EoR/CD 21-cm signal. Some radiation background analyses, like those for the CMB, are very mature. However, the overlap with the 21-cm signal is small, given that backgrounds are integrated over all redshifts, while the 21-cm signal loses many on-sky modes to foregrounds.
2. Resolved galaxies: as shown in Figure 1, galaxy maps would be obvious choices for cross-correlation with the 21-cm signal. Mapping galaxies, at least at lower redshifts, is also a mature field. However, the need for accurate redshift determinations, a reasonably wide field of view, and a sufficient number density of $z > 5.5$ objects, makes it challenging to design wide and deep surveys. Narrow-band dropouts, grism, or spectroscopic follow-up are the most promising options (for example, Wiersma et al., 2013; Vrbanec et al., 2020; Sobacchi, Mesinger & Greig, 2016; Hutter et al., 2017; Moriwaki et al., 2019; Kubota et al., 2020; Heneka & Mesinger, 2020; Hutter et al., 2023; La Plante et al., 2023).
3. Intensity mapping (i.e. unresolved galaxies): other lines such as [CII], [OIII], CO, Lyman-α are mostly sourced by the combined emission from unresolved galaxies. In principle, line intensity maps (LIMs) could have wide fields and good redshift localisation, thus providing an excellent complementary probe to 21-cm observations (for example, Kovetz et al., 2017). However, LIMs have not yet been made at high redshifts, and many line luminosities are expected to be faint (for example, Crites et al., 2014; Yue et al., 2015; Lagache, Cousin & Chatzikos, 2018; Heneka & Cooray, 2021). A reported cross-correlation between two untested probes might be met with skepticism. However, credible detections of (sub)millimetre lines from high-$z$ galaxies with the Atacama Large Millimeter/submillimeter Array (ALMA) should help demonstrate the viability of LIM at higher redshifts in the near future.
4. Quasar spectra: the Lyman-α forest in quasar spectra provides a 1D skewer through the IGM which could be cross-correlated with the corresponding 21-cm forest. These UV and radio absorption lines in the same background object would provide complimentary information about the intervening IGM. Such a study requires the presence of radio-loud quasars at high redshifts, as well as a sufficiently cold IGM so as to detect the 21-cm forest (Bhagwat et al., 2022).

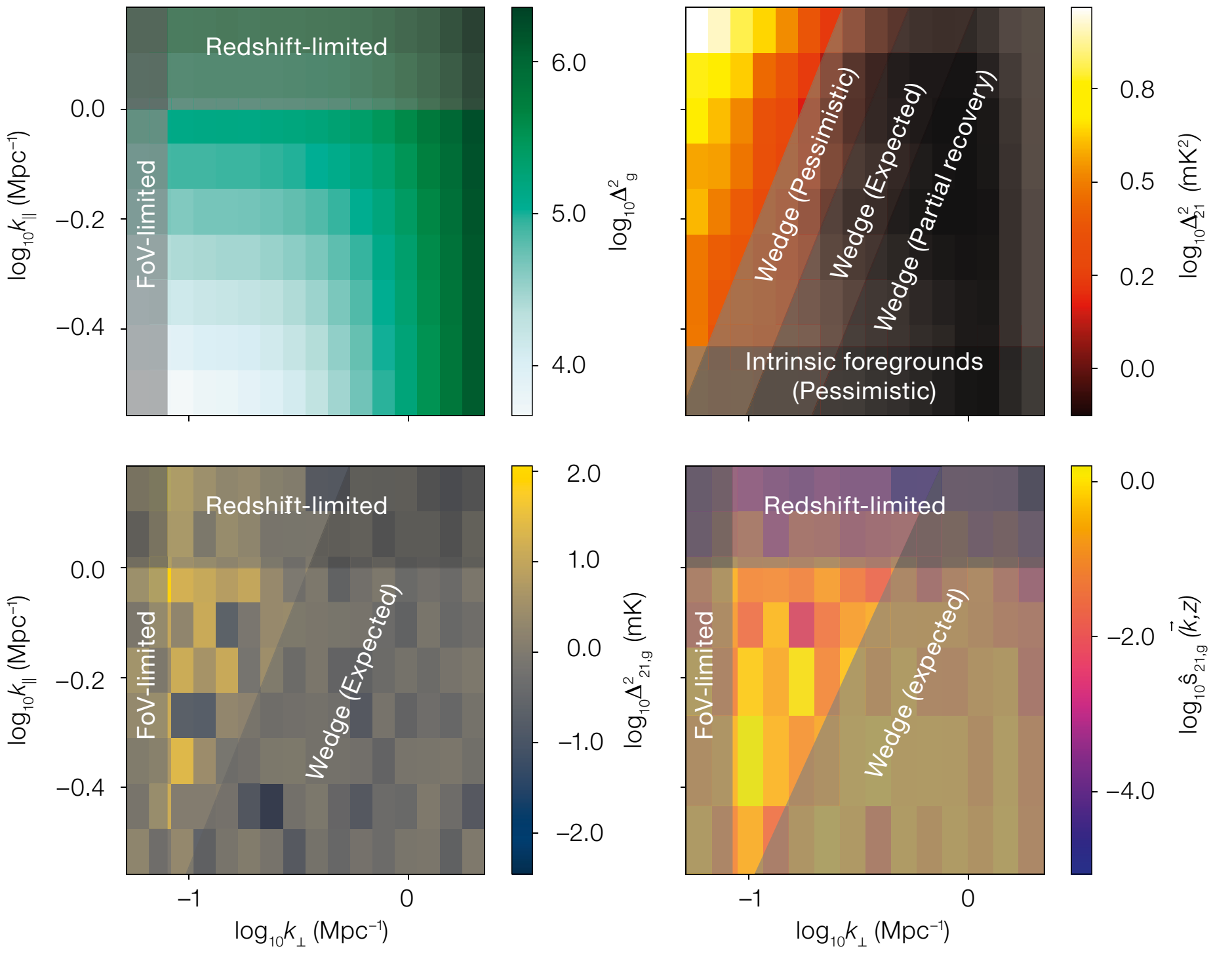

Figure 2. Cylindrical power spectra (PS) of various fields used in the calculation of the galaxy–21-cm cross-power spectrum S/N ratio. Clockwise from top left: galaxy auto PS, 21-cm brightness temperature auto PS, S/N ratio in each bin of the cross-power spectrum, galaxy–21-cm cross-power spectrum. The total S/N of the cross-power spectrum shown in Figure 3 is computed by summing the S/N of each wave-mode bin in quadrature. Figure from Gagnon-Hartman et al. (in preparation).

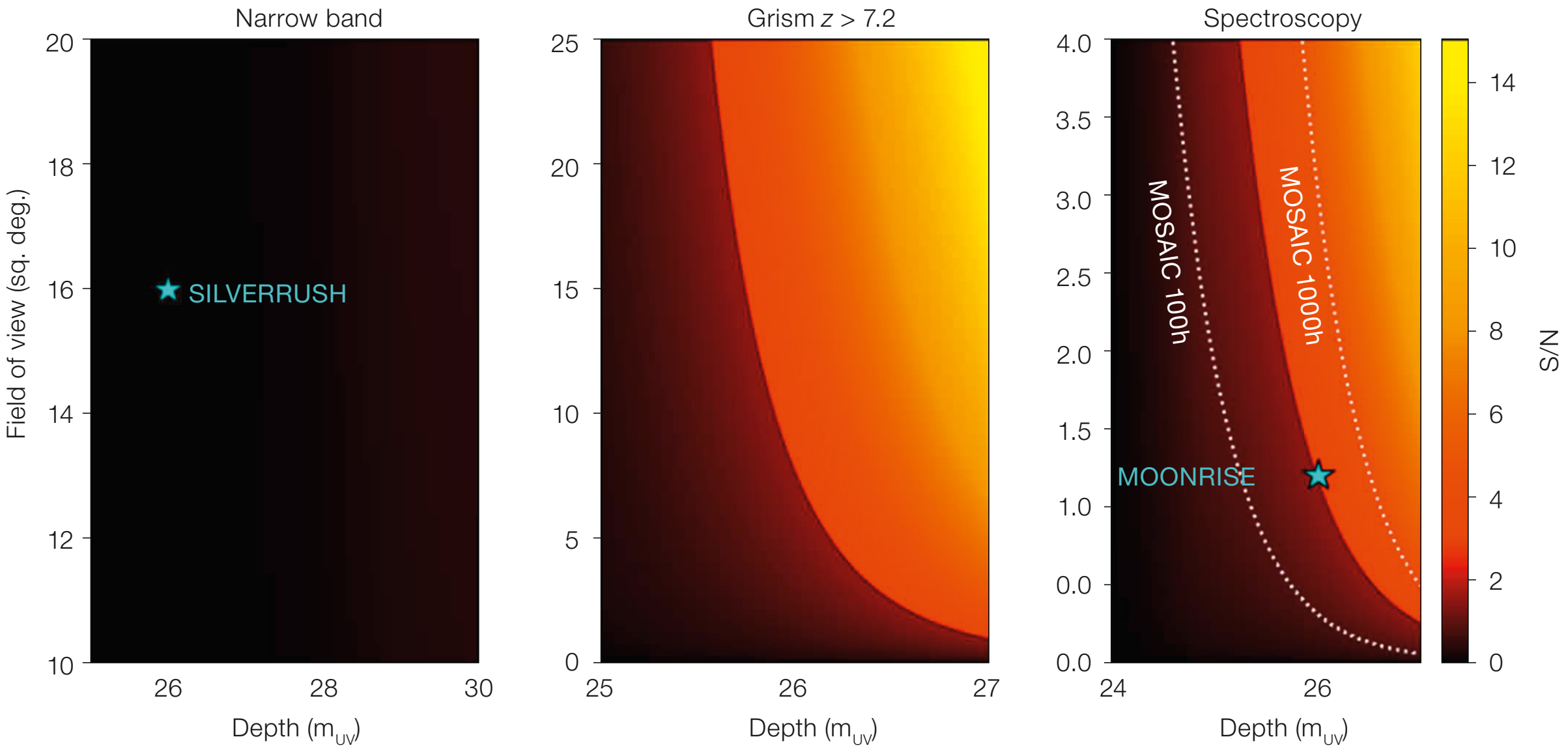

**Figure 3.** Signal-to-noise ratio of the galaxy – 21-cm cross-power spectrum as a function of Lyman-$\alpha$ galaxy survey FoV and apparent UV magnitude limit. Narrow-band dropout, grism, and spectroscopy are shown left to right. The ESO instruments MOSAIC and MOONS have the potential to provide a high-S/N measurement targeting a ~ 1.5-square degree patch of sky. The MOONRISE survey, which targets ~ 1100 sources at $z > 5.5$, will provide a galaxy map sufficient for a S/N ~ 3 detection of the galaxy-21-cm cross-power spectrum. Narrow-band dropout photometry does not provide sufficient redshift precision to allow for a significant detection of the cross-power spectrum. Figure from Gagnon-Hartman et al. (in preparation).

In the following, we focus on specific ESO instruments that have arguably the most potential for a cross-correlation with an EoR/CD observation with SKA-Low. For 21-cm interferometry forecasts with SKA-Low we assume a fiducial, deep (1000-hour) survey and moderate foreground wedge excision from Prelogovic & Mesinger (2023). For 21-cm forest forecasts with SKA-Low, we assume an observation time of 1000 hours and a bandwidth of 10 kHz (see Bhagwat et al., 2022).

## SKAO synergies with ESO facilities

### MOSAIC/MOONS (galaxy Lyman-$\alpha$) — SKA (21-cm intensity mapping)

The Multi-Object Optical and Near-infrared Spectrograph (MOONS) and MOSAIC are multi-object spectrographs currently in development for deployment at the Very Large Telescope (VLT) and ESO's Extremely Large Telescope (ELT), respectively. Both instruments are expected to yield high-precision redshift estimates for thousands of Lyman-$\alpha$ emitters (LAEs) from the EoR. They will cover similar wavelength ranges of 0.76–1.35 and 0.45–1.8 μm, respectively, both at a spectral resolving power $R$ ~ 4000. MOSAIC is expected to have much greater sensitivity since it benefits from the larger dish of the ELT; however, it has a smaller field of view (FoV) and its deployment is delayed relative to that of MOONS. Furthermore, the MOONRISE survey (Maiolino et al., 2020) already has guaranteed observing time on the MOONS instrument to measure the redshifts of about a thousand LAEs at $z > 5.5$. Candidates for future surveys using MOONS or MOSAIC may be provided by photometric or grism surveys with the Nancy Grace Roman Space Telescope or the James Webb Space Telescope.

To measure a cross-correlation of galaxy and 21-cm maps, their effective footprints should overlap. As seen in Figure 2, the FoV of the galaxy survey sets the largest observable on-sky mode in the cross-power spectrum (i.e. the smallest $k_\perp$), and its redshift uncertainty sets the smallest observable line-of-sight mode (i.e. the largest $k_\|$). The 21-cm foreground 'wedge' limits observability in the small $k_\|$ – large $k_\perp$ corner of cylindrical k-space. The amplitude of the cross-power is determined mostly by the depth of the galaxy survey (see La Plante et al., 2023 for a recent forecast for the SKA precursor, Hydrogen Epoch of Reionization Array).

Gagnon-Hartman et al. (in preparation) forecast the potential for cross-correlation between MOONS/MOSAIC galaxy fields and a prospective 21-cm field detected by SKA-Low. They find that a survey comprising roughly 1000 hours of observation time with MOSAIC covering ~ 2 square degrees can detect the 21-cm signal in cross-correlation at a significance level of ~ 6$\sigma$. They also find that the planned MOONRISE survey covering 1.2 square degrees should yield a ~3$\sigma$ detection of the galaxy–21-cm cross-power spectrum.

### ANDES (quasar metal lines) — SKA (21-cm forest)

The ArmazoNes high Dispersion Echelle Spectrograph (ANDES) is the high-resolution spectrograph foreseen for the ELT, whose design phase is ongoing. It will cover the wavelength range 0.4–1.8 μm simultaneously at a resolving power $R$ ~ 100 000, making it ideal for the study of the EoR with quasar absorption spectra (among many other science cases), and in particular using absorption lines from chemical elements heavier than helium (generally called 'metals') which will be fully resolved at this resolution.

Bhagwat et al. (2022) have used hydrodynamic cosmological simulations to test how simultaneous absorption from metals (observed with ANDES) and HI 21-cm (with SKA) in the spectrum of a bright, radio-loud background quasar can complement each other as probes of the underlying gas properties. In Figure 4 we show maps of HI column density overlaid with 21-cm optical depth $\tau_{21cm}$, OI, CII, SiII and FeII column density at various redshifts, resulting from post-processing of the radiation hydrodynamic simulation Aurora (Pawlik et al., 2017). From the figure, the correspondence between a high HI and metal column density is clear.

Such correspondence is better quantified in Figure 5, where we show synthetic spectra from ANDES and SKA-Low. Here we assume a background source with flux of 10 mJy at 150 MHz and a power law index of –1.31. The shaded regions highlight the locations of aligned and cospatial absorbers (ACA), i.e. those in which 21-cm and metal absorption fall in the same window in velocity space and absorption originates from the same underlying gas, while for non-cospatial systems the absorption appears aligned only because of peculiar velocity effects and thus does not effectively probe the same region in space. These observations would probe the physical state of the IGM, including its temperature, metallicity and ionisation state, as well as the nature of nearby sources.

## Conclusions

The first, low-S/N 21-cm EoR detection should come in the near future. The best way of convincing ourselves and the community that this detection is genuine would be to cross-correlate it with a known $z > 5$ signal (that has completely different foregrounds and systematics).

All high-redshift probes ultimately correlate with the matter field, albeit non-trivially in the case of 21 cm owing to its sensitivity to large-scale radiation fields. Thus, there are many candidates for cross-correlations, including: (i) cosmic radiation backgrounds (CMB, CXB, CIB); (ii) (resolved) galaxy maps; (iii) line-intensity maps; and (iv) quasar spectra.

Here we identify promising ESO instruments that will synergise well with SKA-Low EoR surveys: (i) MOONS on the VLT; (ii) MOSAIC on the ELT; (iii) ANDES on the ELT. We show that a Lyman-$\alpha$ galaxy map obtained with MOONS or MOSAIC can be used together with the 21-cm interferometric signal to detect the cross-power spectrum at high S/N. Line-intensity maps obtained with a single-dish telescope with spectroscopic capabilities in the submillimetre (like the proposed concept for the Atacama Large Aperture Submillimeter Telescope [AtLAST; see Klaassen et al., 2020; Mroczkowski et al., 2024]) could also contribute to this. We also demonstrate that correlating Lyman-α and 21-cm absorption as seen in the spectrum of a putative radio-loud high-$z$ quasar can provide complimentary information about the physical state of the IGM.

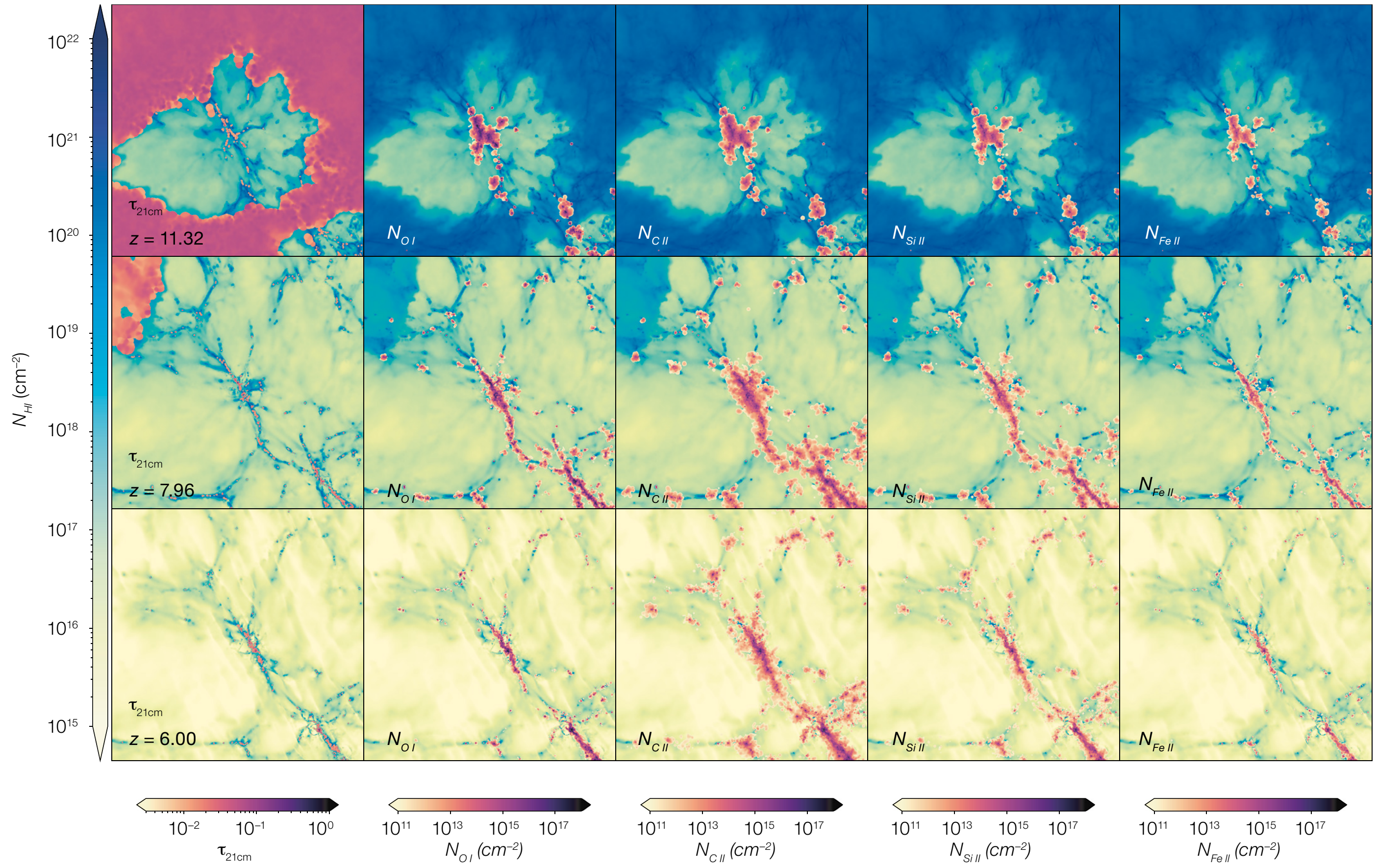

**Figure 4.** Maps of HI column density overlaid with (from left to right): 21-cm optical depth $\tau_{21cm}$, OI, CII, SiII and FeII column density at $z = 11.32$ (top panels), 7.96 (middle) and 6 (bottom). Each projection is 3 $h^{-1}$ c Mpc on a side and 1 $h^{-1}$ c Mpc deep. Figure from Bhagwat et al. (2022).

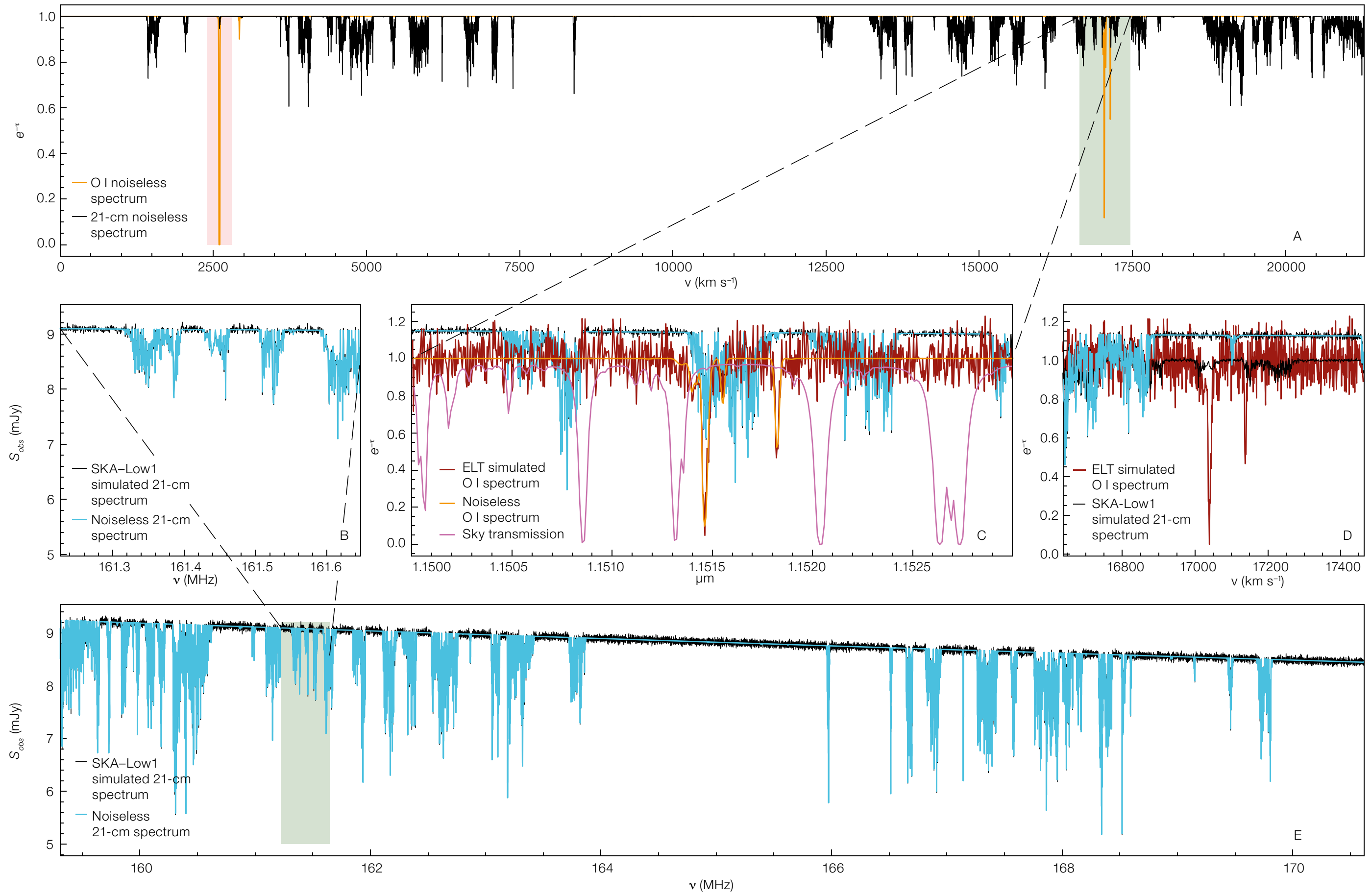

Figure 5. Panel A: synthetic (noiseless) spectra at $z$ = 7.96 showing the 21-cm forest (black lines) and the O I $\lambda$1302 absorption spectra (orange). Shaded regions (pink and green) highlight the location of ACA, but their extent does not coincide with that of the corresponding absorber. The green shaded regions correspond to the same absorber in panels A and E. Panel B: zoom into the portion of the 21-cm forest spectrum with the ACA shown in panels A and E. The blue line is the noiseless spectrum ($e^{-\tau_{21cm}}$), while the black one is the simulated spectrum ($S_{obs} = S_0\, e^{-\tau_{21cm}}$) for a background radio-loud source of flux ($S_0$) of 10 mJy at 150 MHz. Panel C: zoom into the portion of the O I spectrum of Panel A with the ACA. The orange line shows the noiseless spectrum, while the red one shows the simulated spectrum as observed by the ELT–ANDES instrument assuming an exposure time of 10 hours, two readouts per hour and spectral resolution $R = 10^5$ for a $m_{AB}$ = 21.0 mag source. The purple line shows sky transmission at these wavelengths. Panel D: simulated normalised spectra as observed by the ELT and SKA-Low for O I and 21 cm overlayed in velocity space showing the green shaded region from Panel A. Panel E: full extent of the simulated 21-cm forest. The blue line shows the noiseless spectrum (as in panel A) multiplied with the radio continuum of the background source, while the black line shows the simulated spectrum as observed by SKA-Low. Figure from Bhagwat et al. (2022).

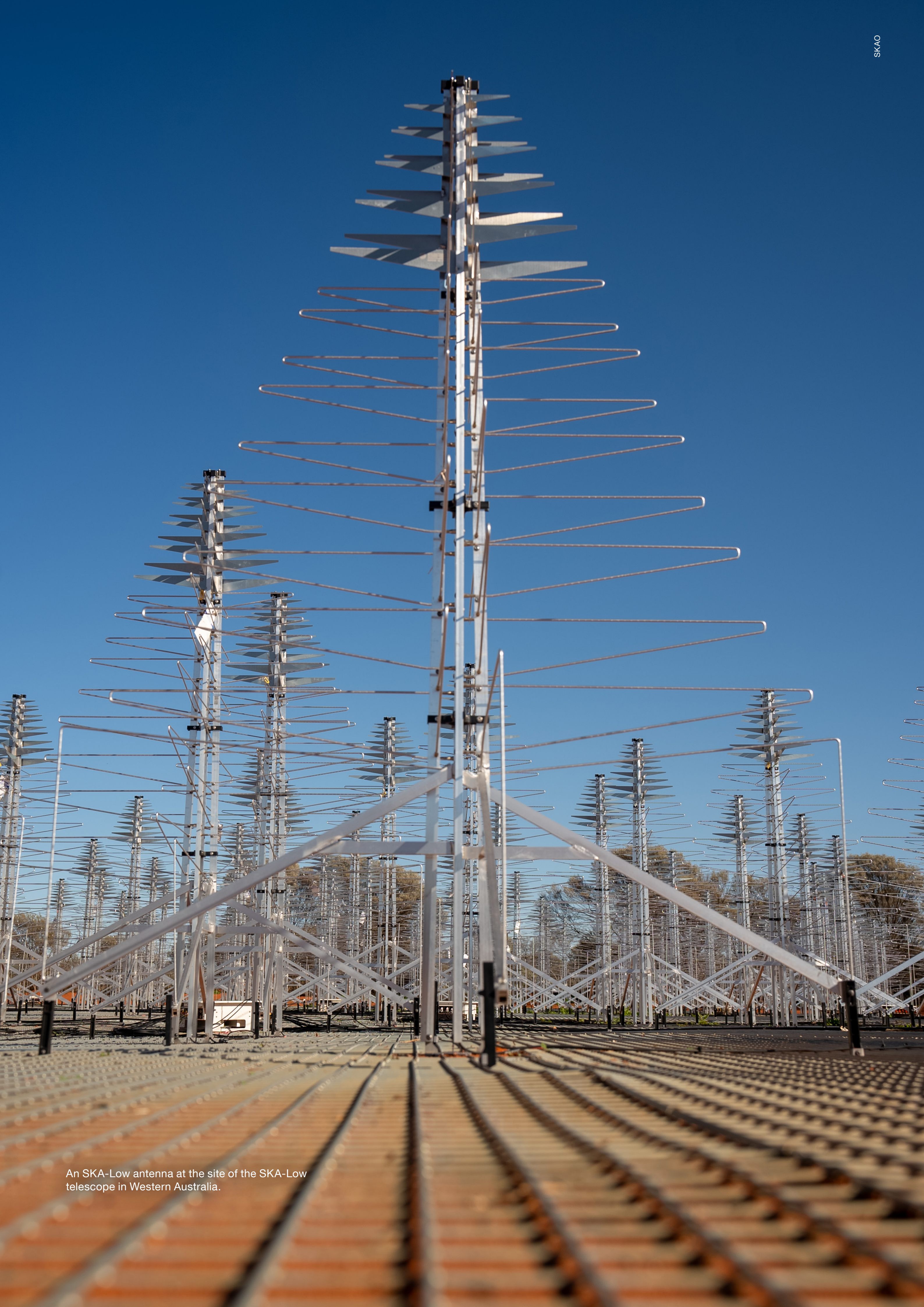

SKAO

An SKA-Low antenna at the site of the SKA-Low telescope in Western Australia.